\documentclass[onecolumn,amsmath,amssymb]{qm2008_abs}

\usepackage{graphicx}
\usepackage{dcolumn}
\usepackage{bm}

\newcommand{\sqrtsnn}{\sqrt{s_{_{NN}}}}

\begin{document}

\title{{\Large Cosmic-ray Monte Carlo predictions for forward particle production 
in p-p, p-Pb, and Pb-Pb collisions at the LHC }}

\bigskip
\bigskip
\author{\large David d'Enterria$^{1,}$\footnote{Support from 6th EU Framework Programme MEIF-CT-2005-025073 is acknowledged.},
Ralph Engel$^{2}$, \underline{Thomas McCauley}$^{3,}$\footnote{Corresponding author: thomas.mccauley@cern.ch}, and Tanguy Pierog$^{2}$}
\affiliation{$^1$CERN, PH, CH-1211 Geneva 23}
\affiliation{$^2$Forschungszentrum Karlsruhe, Postfach 3640, D - 76021 Karlsruhe}
\affiliation{$^3$Northeastern University, 111 Dana, Boston, MA 02115}
\bigskip
\bigskip

\begin{abstract}
\leftskip1.0cm
\rightskip1.0cm
We present and compare the predictions of various cosmic-ray Monte Carlo models 
for the energy  ($dE/d\eta$) and particle ($dN/d\eta$) flows in p-p, p-Pb and Pb-Pb 
collisions at $\sqrtsnn =$~14, 8.8, and 5.5 TeV respectively, in the range covered 
by forward LHC detectors like CASTOR or TOTEM (5.2$<|\eta|<$6.6)
and ZDC or LHCf ($|\eta|\gtrsim$8.1 for neutrals).
\end{abstract}

\maketitle

\section{Introduction}
\begin{figure}[htbp]
\centering{\includegraphics[width=7.cm]{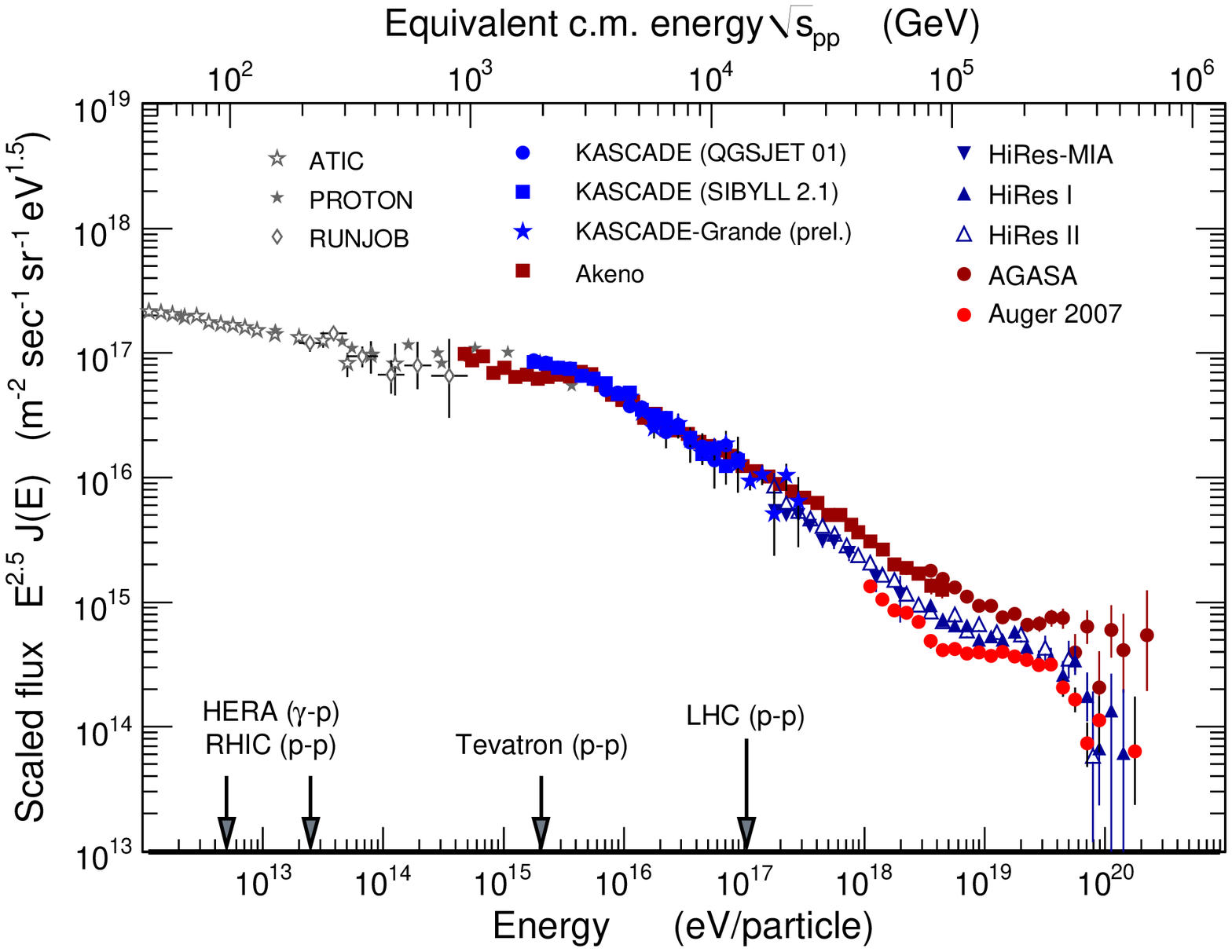}
\includegraphics[width=7.cm]{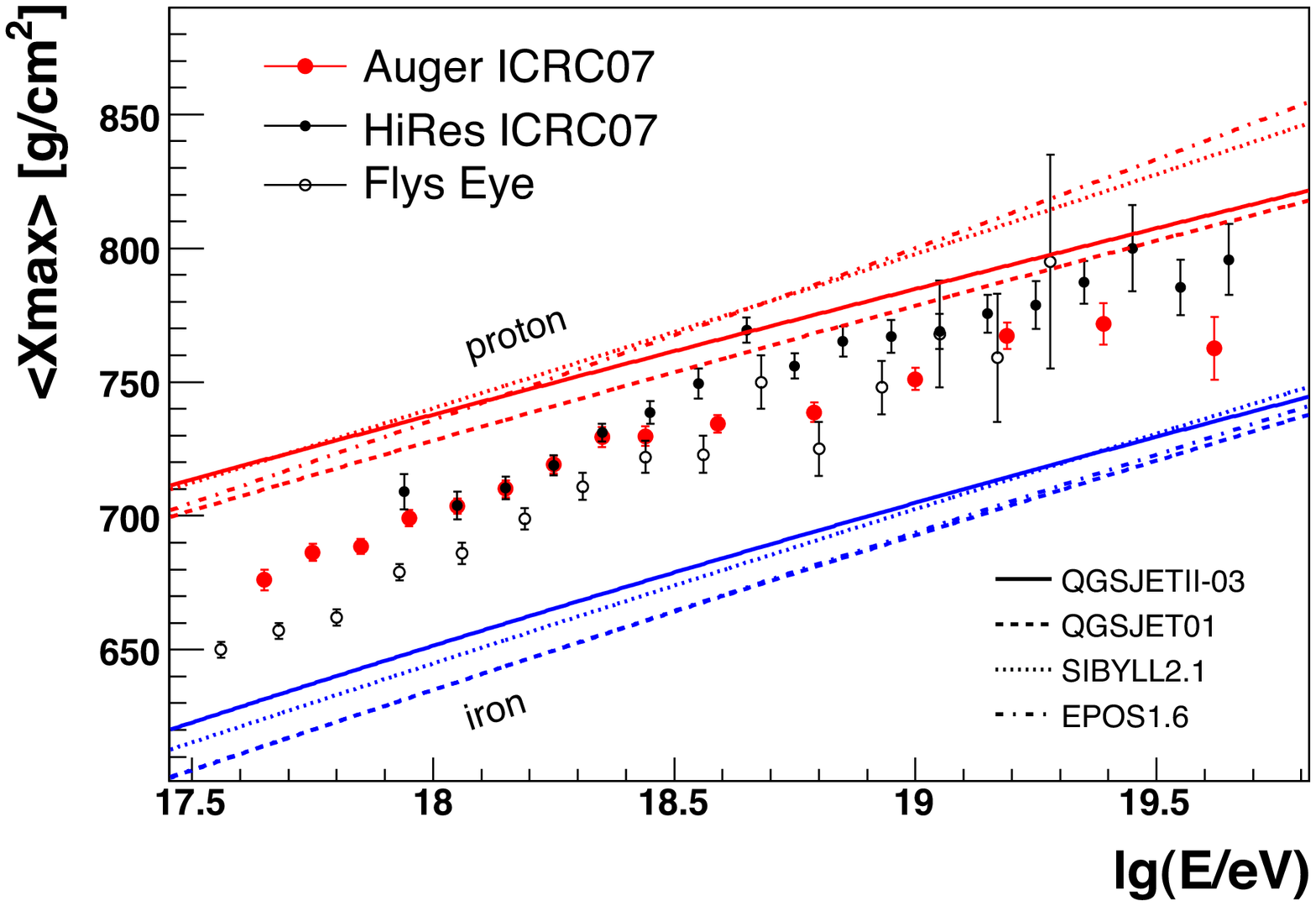}}
\caption{Left: Measured cosmic ray energy spectrum compared to the center-of-mass energy of various
hadron colliders. Right: Measurements of CRs shower maximum in the atmosphere as a function of 
primary energy for several experiments along with MC predictions for proton and iron primaries.}
\label{fig:CRs}
\end{figure}

The origin and nature of cosmic rays (CRs) with energies between $10^{15}$~eV
and the Greisen-Zatsepin-Kuzmin (GZK) cutoff at about $10^{20}$\,eV, recently measured by 
the HiRes~\cite{hires} and Auger~\cite{auger}  experiments, remains a central open question 
in high-energy astrophysics. One key to solving this question is the determination of the 
elemental composition of cosmic rays in this energy range. The candidate particles, ranging from 
protons to nuclei as massive as iron, generate ``extended air-showers'' (EAS) in 
interactions with air nuclei when entering the Earth's atmosphere.
Due to their low observed flux (Fig.~\ref{fig:CRs}, left), only indirect (yet complementary)
measurements are possible 
using the atmosphere as ``calorimeter''. The first method relies on measuring the fluorescence 
light emitted by air molecules excited by the cascade of secondaries. 
The second one relies on the use of either scintillators or water \v{C}erenkov tanks to sample the shower at ground. 
Recent Auger results are consistent with the hypothesis that the highest energy 
CRs are protons (source correlated with Active Galactic Nuclei)~\cite{auger2}. However, fluorescence-based 
measurements of the shower maximum as a function of primary energy 
favour a mixed composition (Fig.~\ref{fig:CRs}, right). 

The determination of the primary energy (from the surface detectors alone) and mass (from either method)
relies on hadronic Monte Carlo (MC) codes which describe the interactions of the primary cosmic-ray 
in the upper atmosphere. The bulk of the primary particle production is dominated by forward and soft 
QCD interactions, modeled commonly in Regge-Gribov-based~\cite{Gribov:1968fc} approaches with 
parameters constrained by the existing collider data ($E_{lab}\lesssim$~10$^{15}$~eV).
When extrapolated to energies around the GZK-cutoff, the current MCs predict energy and multiplicity flows 
differing by factors as large as three, with significant inconsistencies in the forward region. 
The coming energy frontier for hadron collisions will be reached by the Large Hadron Collider (LHC). 
The LHC will boast a full compliment of detectors in almost the full range of pseudorapidity (Fig.~\ref{fig:LHC}).
Measurement of forward particle production in p-p, p-Pb, and Pb-Pb collisions at LHC energies 
($E_{lab} \approx 10^{17}$ eV) will thus provide strong constraints on these models and allow 
for more reliable determinations of the CR energy and composition at the highest energies.

\section{Forward detectors at the LHC}

\begin{figure}[htpb]
\centering\includegraphics[width=0.6\textwidth,height=6.cm]{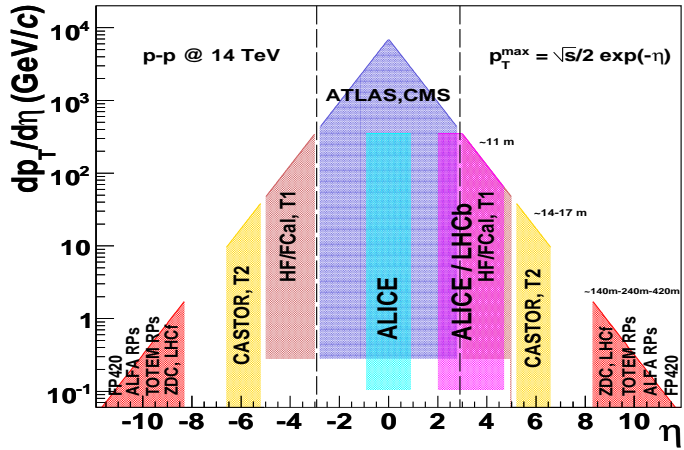}
\caption{Approximate $p_T$-$\eta$ coverage of current (and proposed) detectors at the LHC~\cite{dde}.} 
\label{fig:LHC}
\end{figure}

All LHC experiments feature 
a detection coverage at all rapidities without parallel compared to previous colliders (Fig.~\ref{fig:LHC}). 
ATLAS~\cite{atlas} and CMS~\cite{cms,Albrow:2006xt} not only cover the largest $p_T$-$\eta$ ranges at
mid-rapidity for hadrons, $e^\pm$, $\gamma$ and $\mu$'s, but they also feature extended instrumentation 
at distances far away from the interaction point (IP). Forward calorimetry is available at $\pm$11 m 
(the FCal and HF hadronic calorimeters), at $\pm$14~m (CMS CASTOR sampling calorimeter)~\cite{castor}, 
and at $\pm$140~m (the Zero-Degree-Calorimeters, ZDCs)~\cite{zdc_atlas,Grachov:2006ke}.
In addition, ATLAS has Roman Pots (RPs) at $\pm$220,240~m, and there are advanced plans 
to install a new proton-tagger system at 420~m (FP420) from both the ATLAS and CMS IPs~\cite{fp420}.
Both ALICE~\cite{alice} and LHCb~\cite{lhcb} have forward muon spectrometers 
in regions, 2~$\lesssim~\eta~\lesssim~5$, not covered by ATLAS or CMS. 
The TOTEM experiment~\cite{totem}, sharing IP5 with CMS, features two types 
of trackers (T1 and T2 telescopes) covering $3.1<|\eta|<4.7$ and $5.2<|\eta|<6.5$ 
respectively, plus proton-taggers (Roman Pots) at $\pm$147 and $\pm$220 m.
Last but not least, the LHCf experiment~\cite{lhcf} has installed scintillator/silicon 
calorimeters in the same region of the ATLAS ZDCs, $\pm$140 m away from IP1.

\section{Model predictions}

The bulk of particle production in high-energy hadronic collisions can still not 
be calculated within first-principles QCD. Phenomenological models based on 
general principles such as unitarity and analyticity are often combined 
with perturbative QCD predictions for high-$p_{T}$ processes to obtain an
almost complete description of the final states. Such ``QCD-inspired'' models  
-- like DPMJET~\cite{dpmjet3}, EPOS~\cite{epos}, neXus~\cite{nexus}, 
QGSJET 01 and II~\cite{qgsjet01,qgsjetII}, and SIBYLL~\cite{sibyll}
frequently used for simulating cosmic ray cascades --  use an unitarization procedure 
to reconstruct amplitudes for exclusive processes and to determine the total and 
elastic cross sections. Soft processes are described within Gribov's Reggeon
theory~\cite{Gribov:1968fc} and hadrons are produced mainly in the
fragmentation of color strings. Whereas this Regge-Gribov approximation is
applied to hadrons as interacting objects in the case of DPMJET, QGSJET and SIBYLL, 
it is extended to include partonic constituents in EPOS and neXus. 
Other important differences among models are the approximations of high 
parton density effects at small values of $x=p_{\mbox{\tiny{\it parton}}}/p_{\mbox{\tiny{\it proton}}}$
and the treatment of the hadronic remnants in collisions.
SIBYLL and DPMJET use an energy-dependent transverse momentum cutoff 
for minijet production based on the geometric criterion that there cannot
be more gluons in a hadron than would fit in the given transverse
space~\cite{Bopp:1994cg}. 
QGSJET II implements enhanced Pomeron-Pomeron interactions which are 
equivalent to impact-parameter and density-dependent parton saturation 
for soft processes~\cite{Ostapchenko:2005nj}, achieving a good description 
of RHIC Au-Au data.
EPOS, on the other hand, uses the wealth of RHIC data to parametrize 
the low-$x$ behaviour of the parton densities and the string 
fragmentation parameters~\cite{Werner:2005jf}, reaching a good 
reproduction of a wider range of experimental results.\\

\begin{figure}[htpb]
\centering{\includegraphics[width=0.49\textwidth,height=5.25cm]{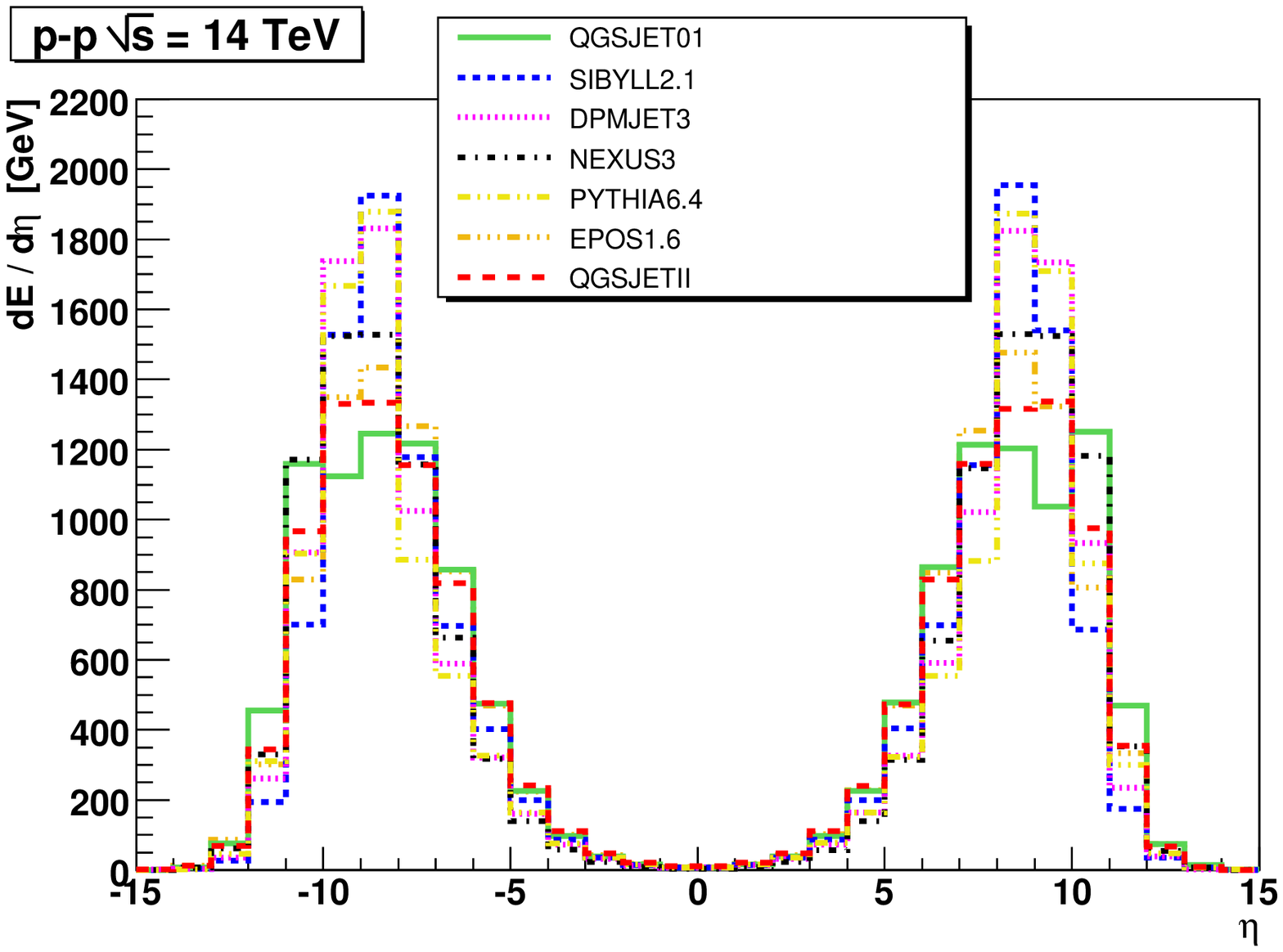}
\includegraphics[width=0.49\textwidth,height=5.25cm]{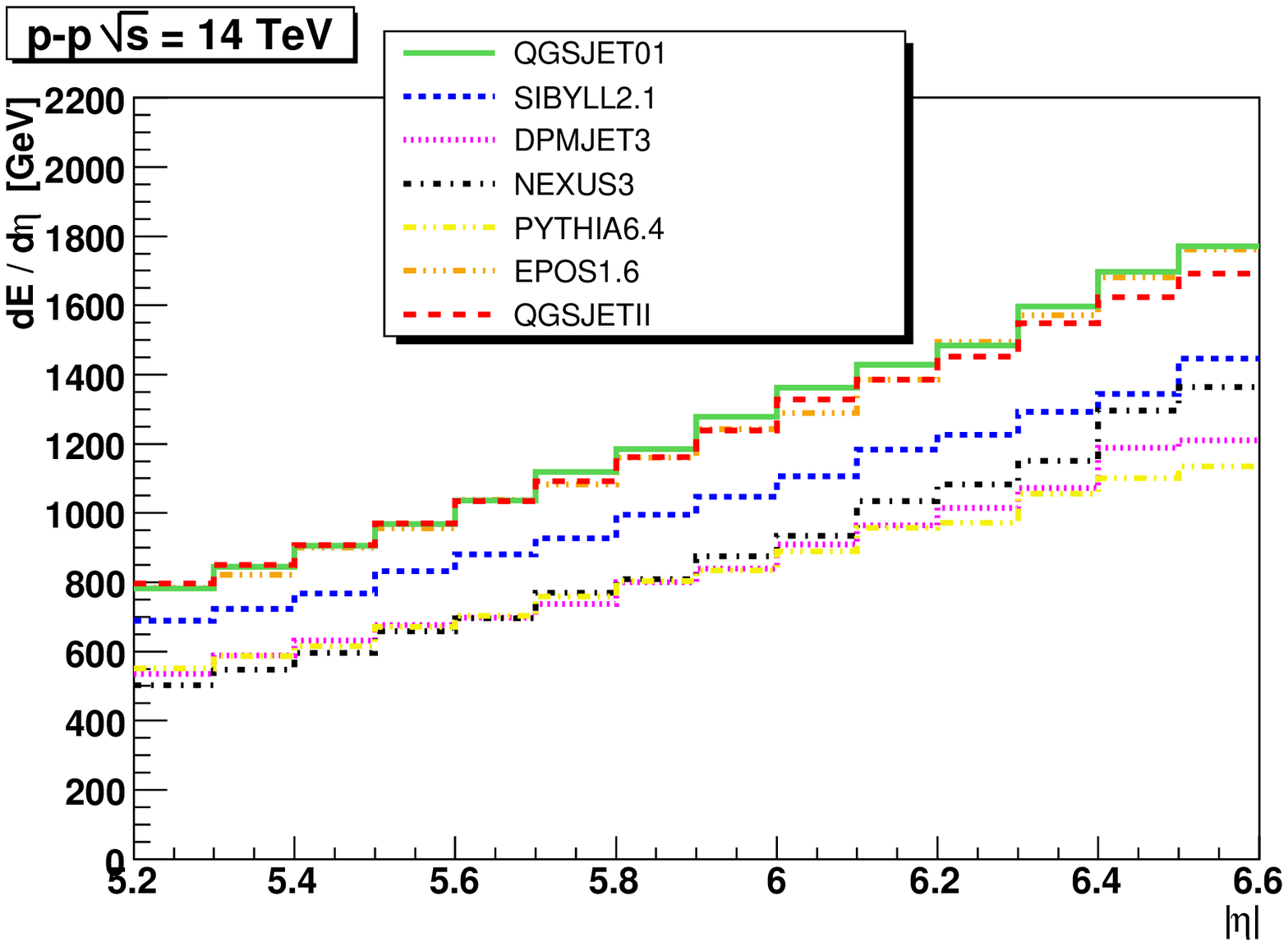}}
\caption{MC predictions for p-p collisions at 14 TeV: energy flow ($dE/d\eta$, left) 
and energy deposited in the CASTOR/T2 acceptance ($dE/d\eta|_{|\eta|=5.2-6.6}$, right).} 
\label{fig:pp}
\end{figure}

\begin{figure}[htpb]
\centering{
\includegraphics[width=0.49\textwidth,height=4.5cm]{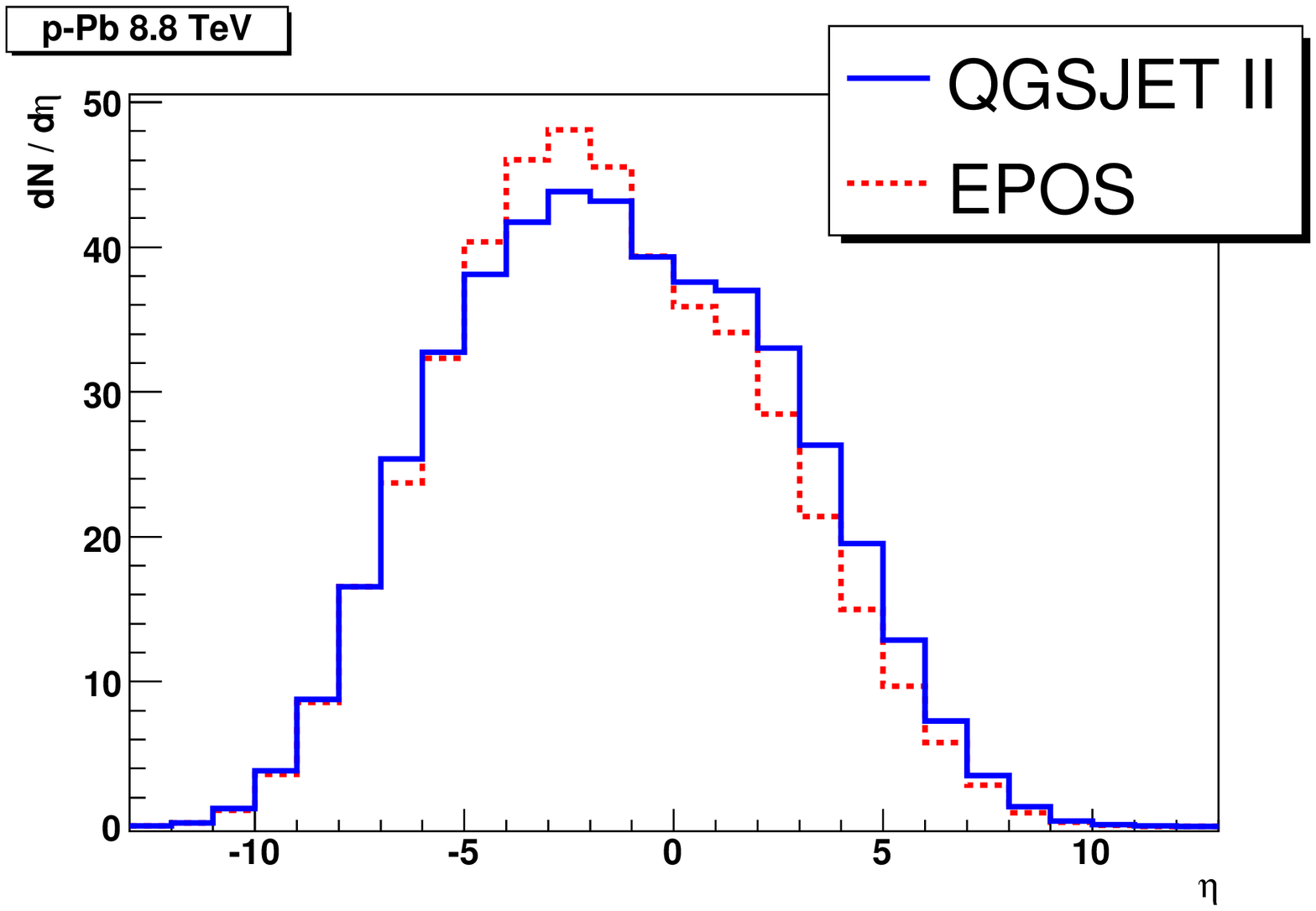}
\includegraphics[width=0.49\textwidth,height=4.5cm]{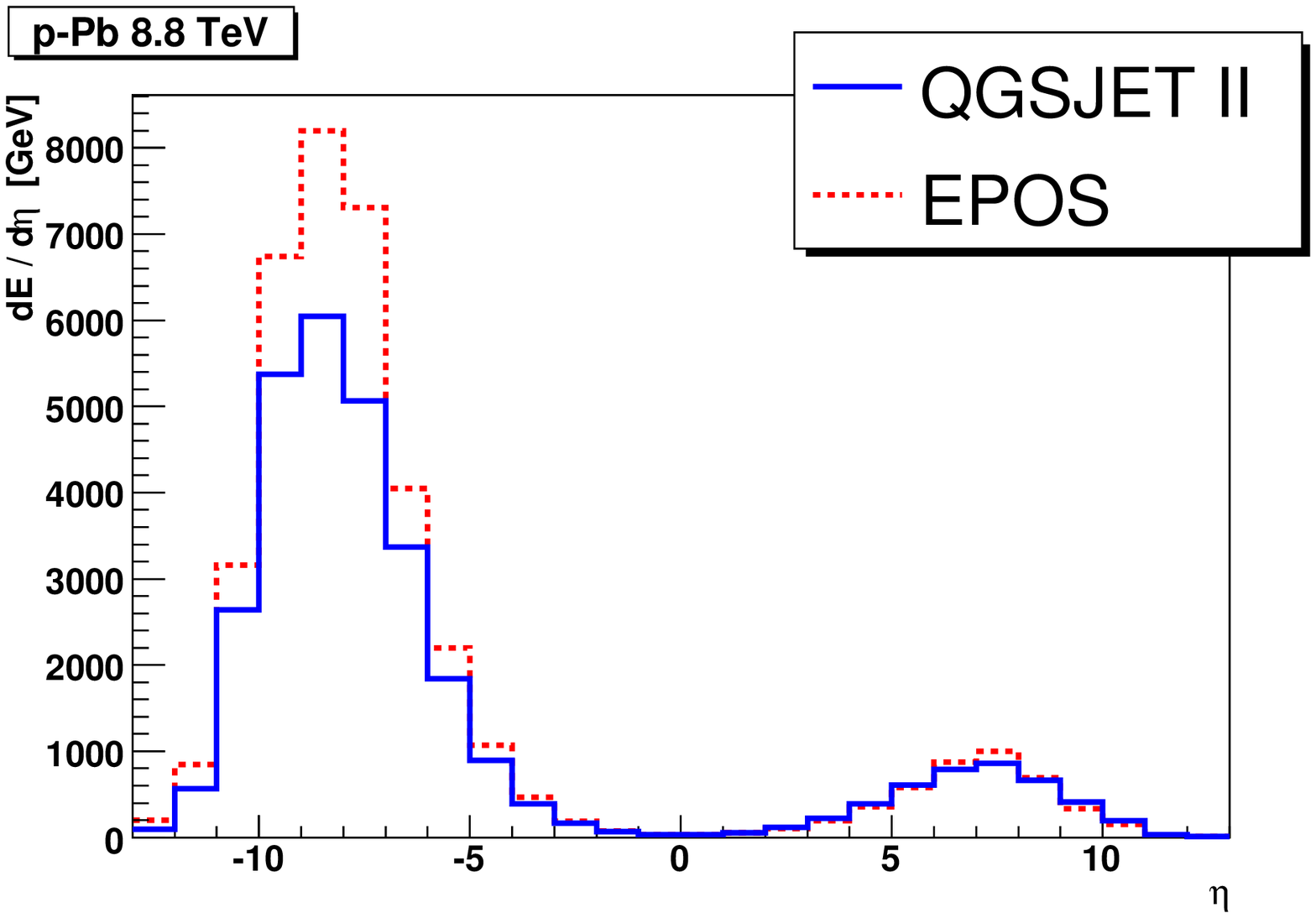}
\includegraphics[width=0.49\textwidth,height=4.5cm]{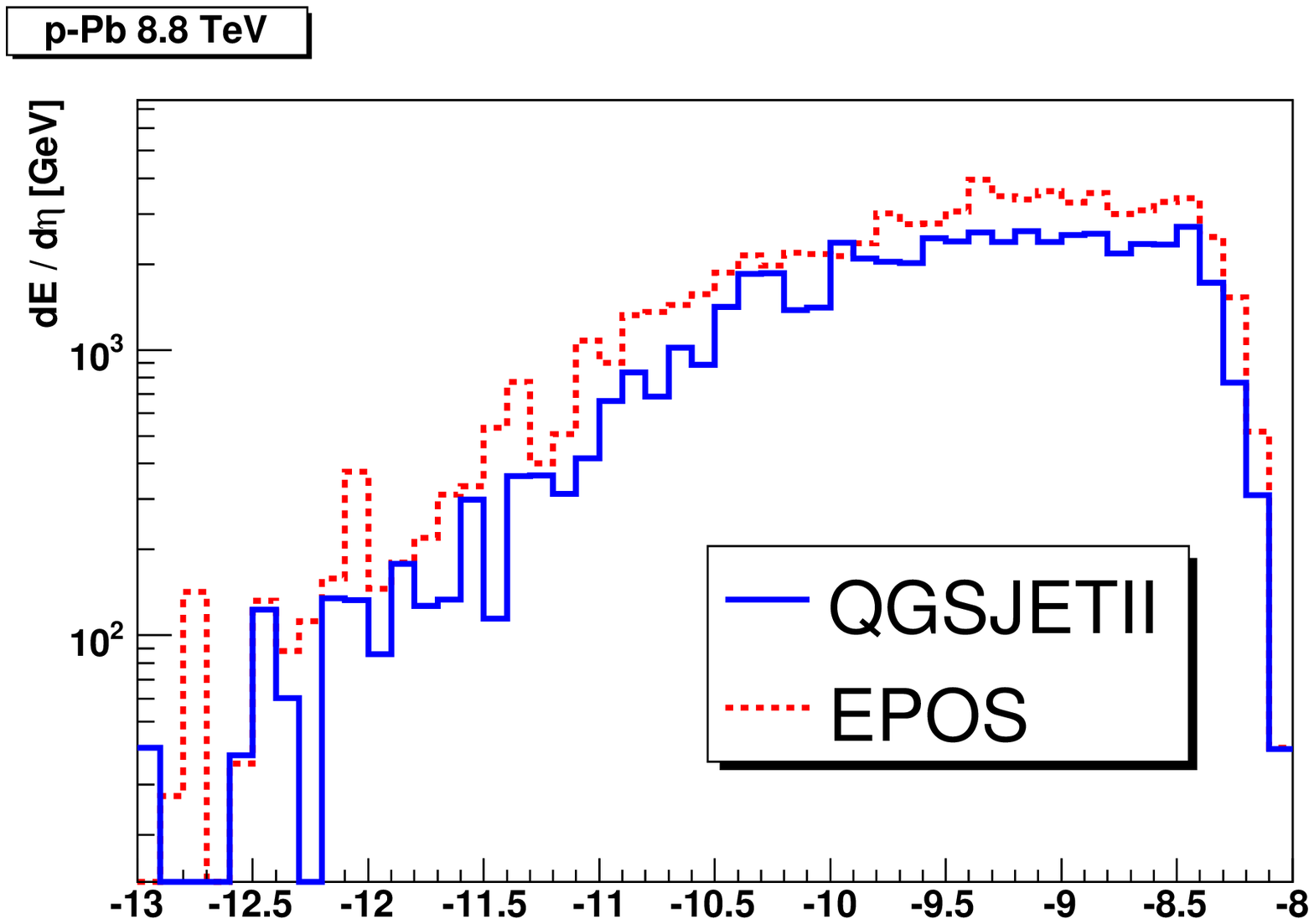}
\includegraphics[width=0.49\textwidth,height=4.5cm]{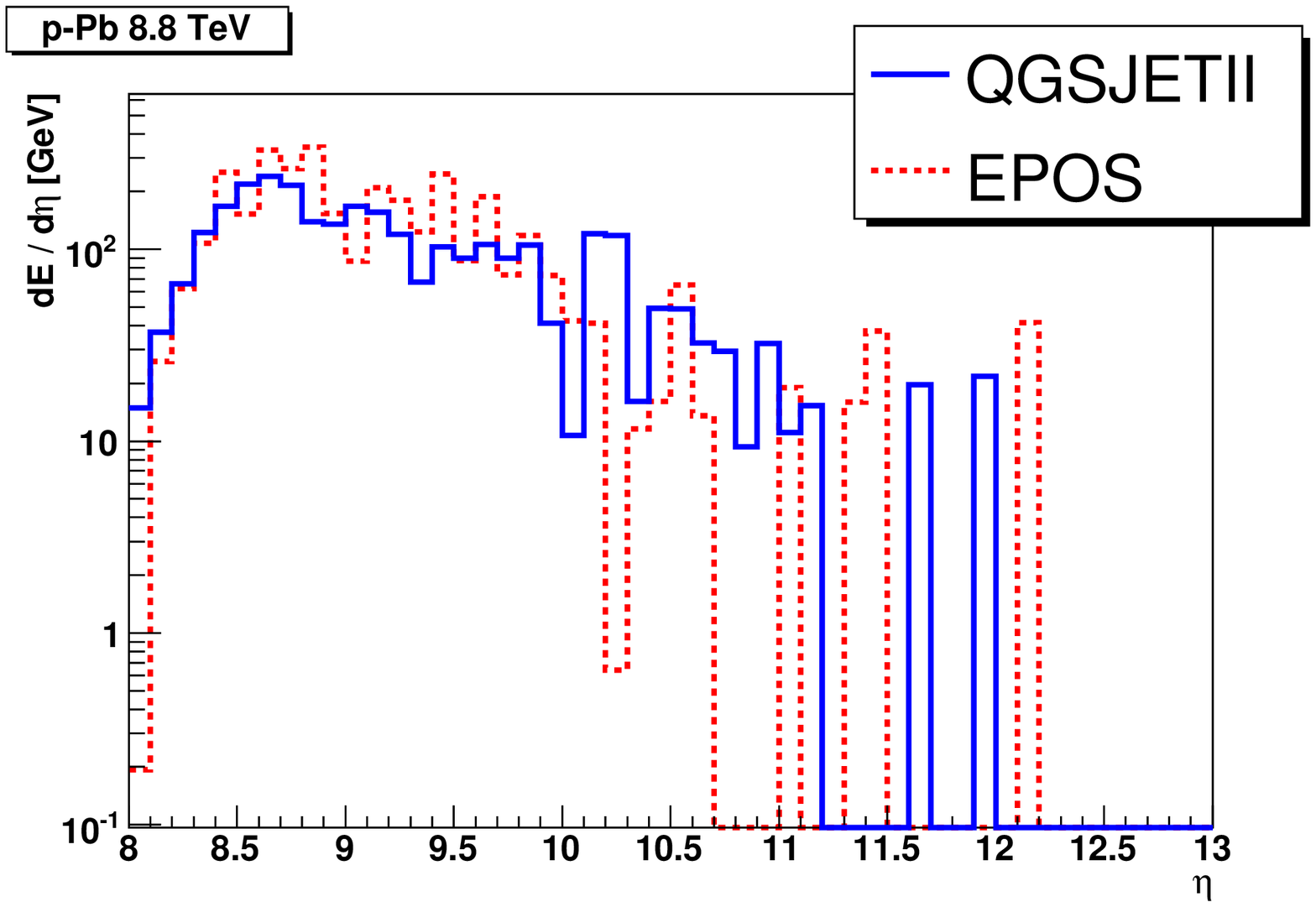}}
\caption{MC predictions for p-Pb collisions at 8.8~TeV: Inclusive particle multiplicity ($dN/d\eta$) and energy 
($dE/d\eta$) densities (top), neutral energy flow at rapidities covered by the ZDC/LHCf detectors (bottom).} 
\label{fig:pPb}
\end{figure}

\begin{figure}[htpb]
\centering{
\includegraphics[width=0.49\textwidth,height=4.5cm]{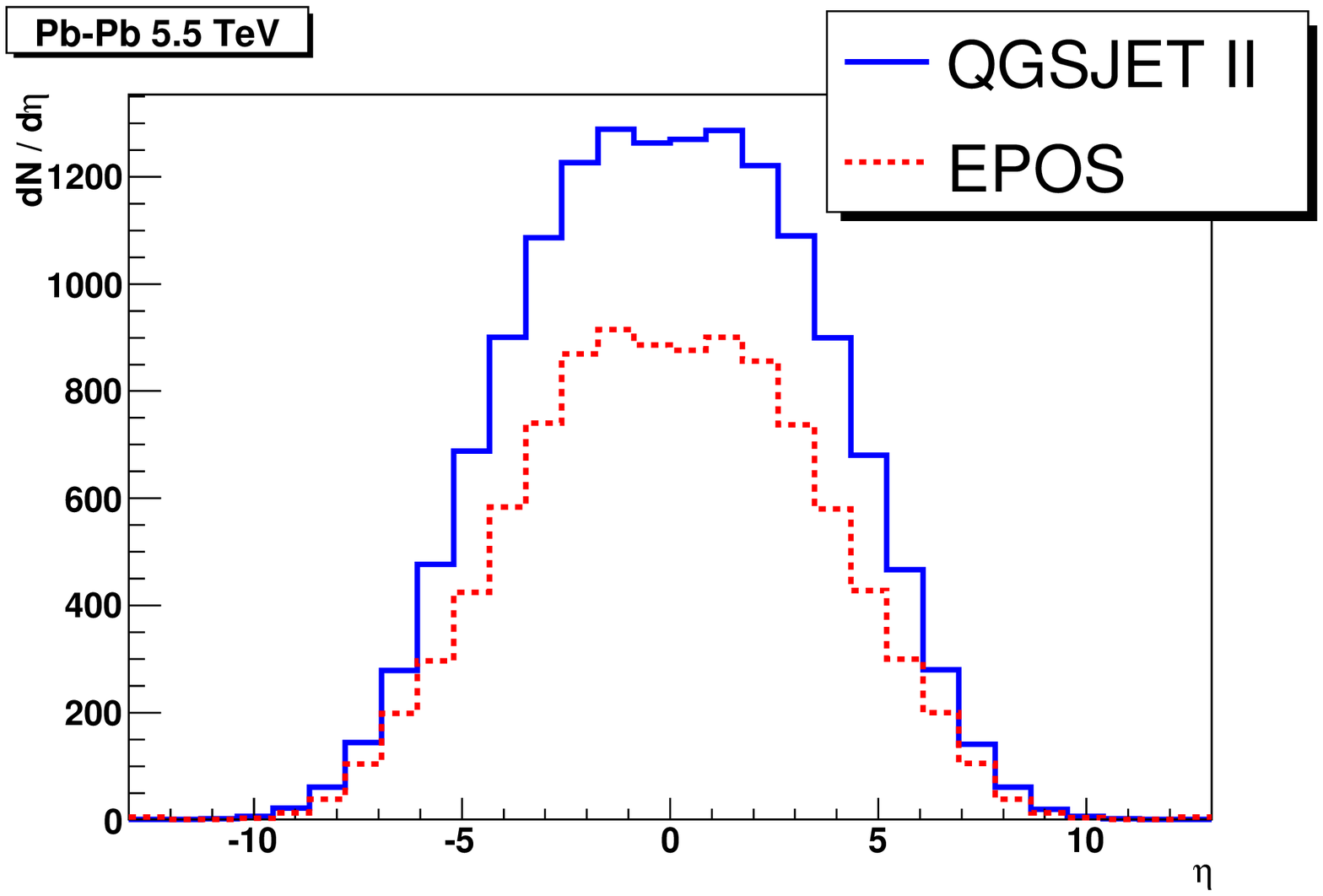}
\includegraphics[width=0.49\textwidth,height=4.5cm]{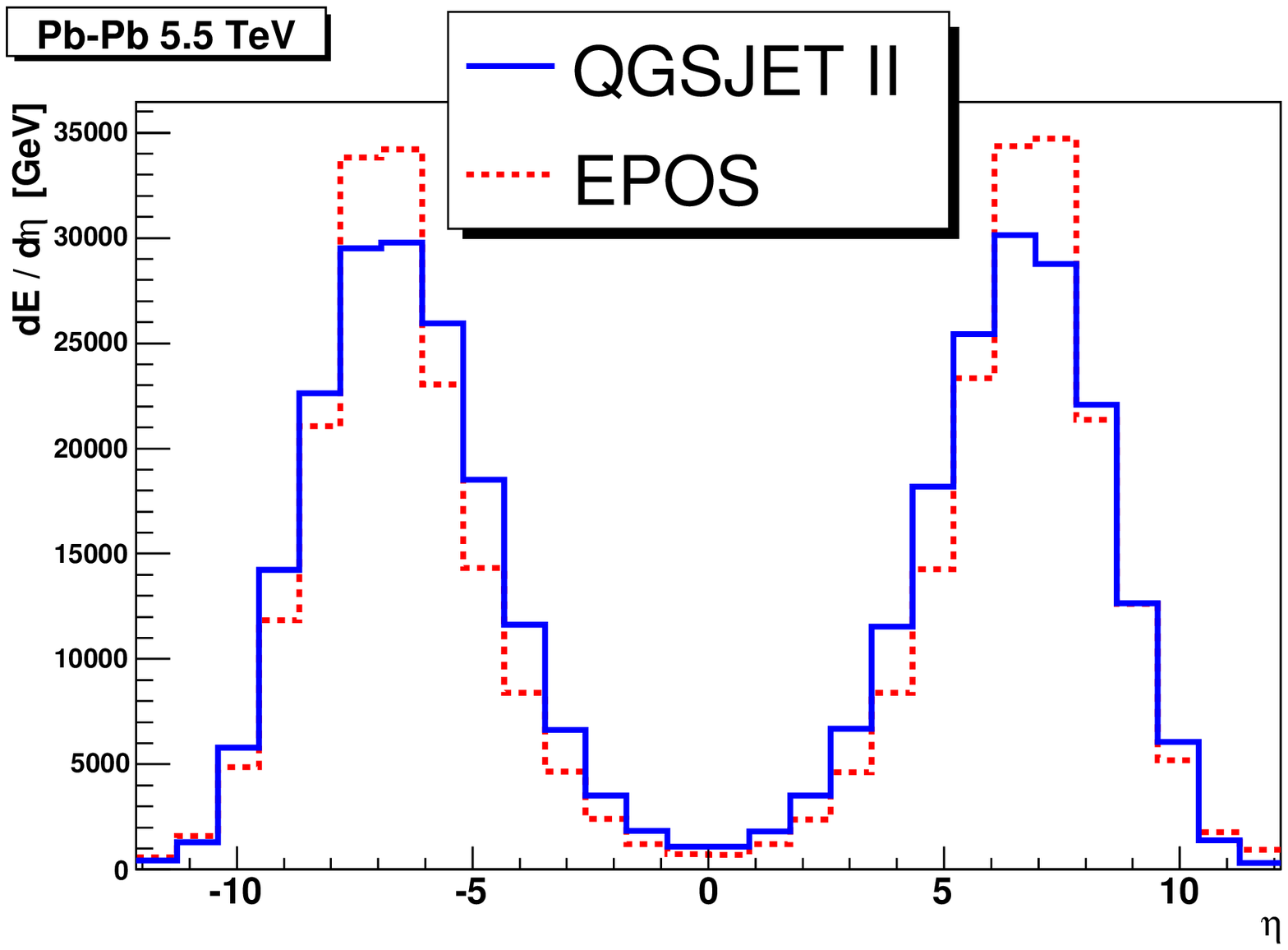}
\includegraphics[width=0.49\textwidth,height=4.5cm]{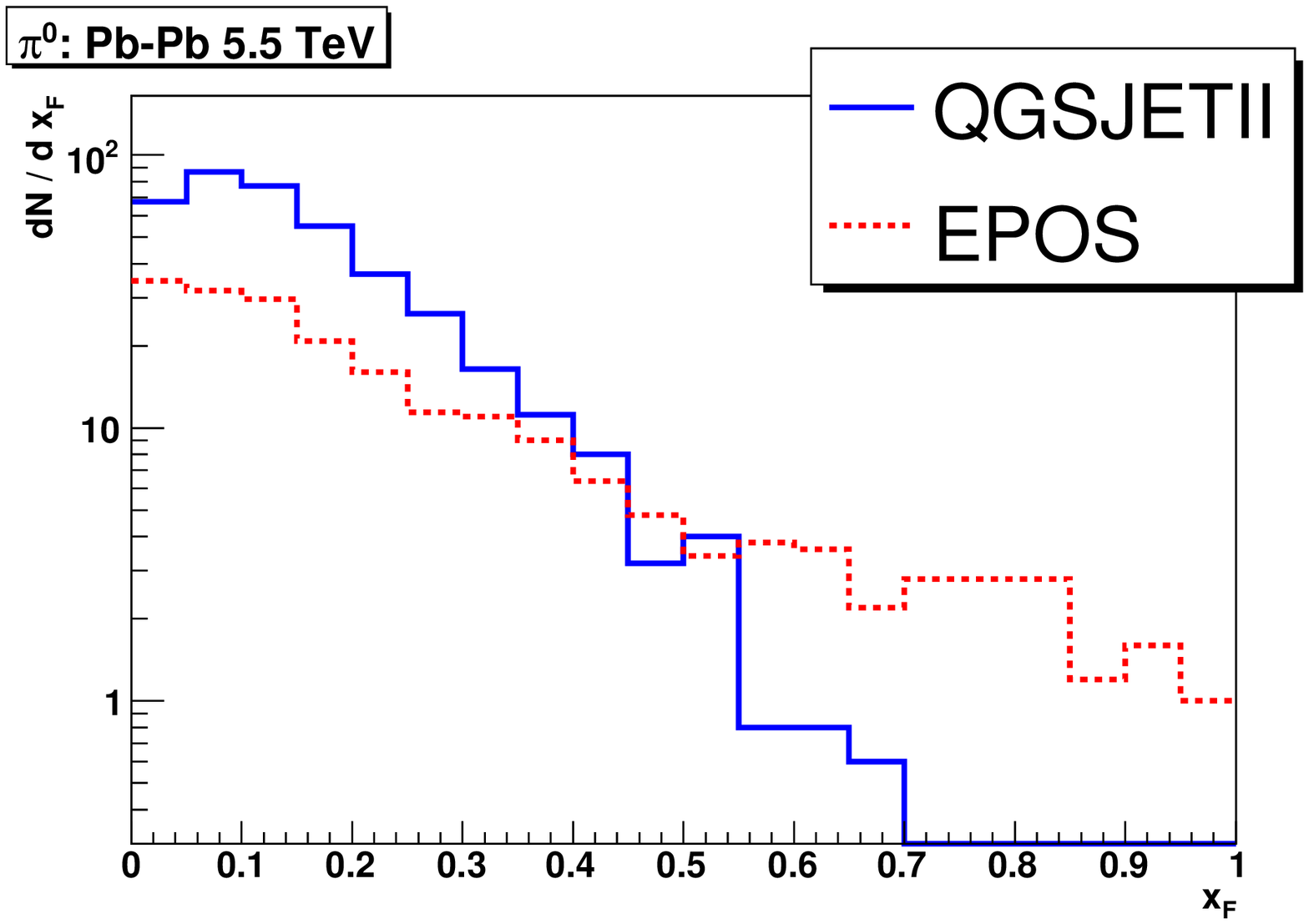}
\includegraphics[width=0.49\textwidth,height=4.5cm]{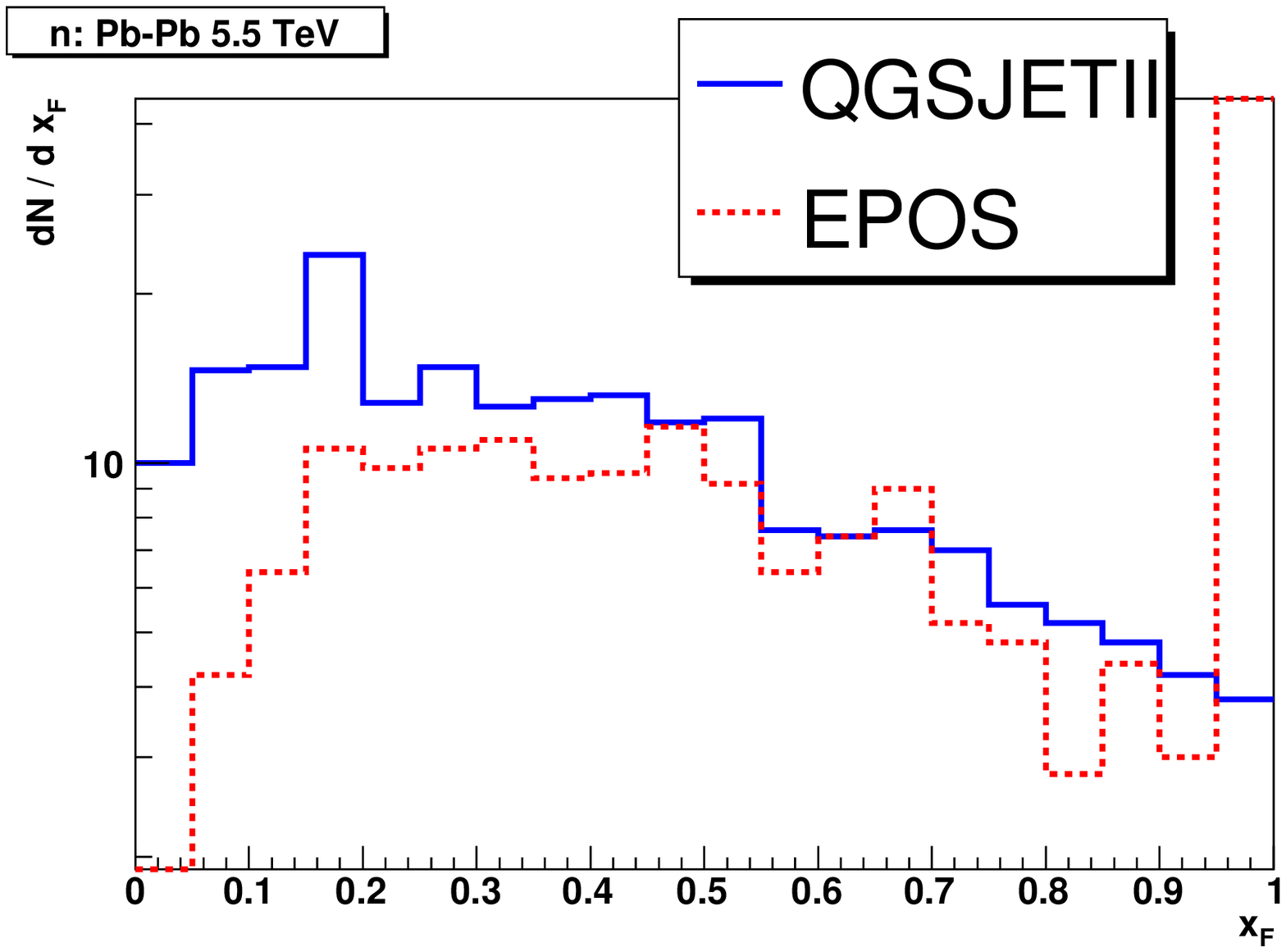}}

\caption{MC predictions for Pb-Pb collisions at 5.5~TeV: Inclusive particle multiplicity ($dN/d\eta$) and energy 
($dE/d\eta$) densities (top), and $x_{F}$ distributions for $\pi^{0}$s and neutrons (bottom).}
\label{fig:PbPb}
\end{figure}

Figures~\ref{fig:pp},~\ref{fig:pPb} and~\ref{fig:PbPb} show for p-p, p-Pb and Pb-Pb respectively, 
the inclusive multiplicity and energy flows predicted by the models (PYTHIA~\cite{pythia} p-p predictions, 
with tunings as used in~\cite{dova}, are also included) for all pseudo-rapidities, as well as the energy deposit in the acceptances
covered by the CASTOR/T2 (5.2$<|\eta|<$6.6) and ZDC/LHCf ($|\eta|\gtrsim$8.1 for neutrals) 
detectors. In some cases, differences as large as 60\% are observed. 
The excellent and complementary detection capabilities of the various experiments at the LHC 
will allow for a (fast) validation and tuning of the MC hadronic models -- already with a first p-p run
with very low integrated luminosity.\\

\end{document}